\documentclass[12pt,a4paper]{article}
\usepackage[a4paper,margin=2.5cm]{geometry}
\usepackage{amsmath,amssymb}
\usepackage{bm}
\usepackage{graphicx}
\usepackage{booktabs}
\usepackage{authblk}
\usepackage{cite}
\usepackage{hyperref}
\hypersetup{colorlinks=true,linkcolor=blue,citecolor=blue,urlcolor=blue,pdftitle={Bogoliubov-de Gennes spectrum, Cherenkov radiation and recoil of solitons with third-order dispersion underlying optical analogue horizons},pdfauthor={Hasan Oguz},pdfkeywords={solitons, third-order dispersion, Cherenkov radiation, soliton recoil, Bogoliubov-de Gennes spectrum, optical event horizon, analogue gravity}}

\newcommand{\kg}{\kappa_g}
\newcommand{\dtres}{\delta_3}
\newcommand{\cmatch}{c_{\mathrm{match}}}
\newcommand{\numinus}{\nu_-}
\providecommand{\openone}{\leavevmode\hbox{\small1\normalsize\kern-.33em1}}

\title{Bogoliubov-de Gennes spectrum, Cherenkov radiation and recoil of solitons with third-order dispersion underlying optical analogue horizons}
\author[1,2]{Hasan Oguz}
\affil[1]{Department of Computer Technologies, Vocational School, Istanbul Okan University, 34959 Tuzla, Istanbul, T\"urkiye}
\affil[2]{Material Physics Simulation Laboratory, Science Faculty, Pamukkale University, 20160 Pamukkale, Denizli, T\"urkiye}
\affil[ ]{E-mail: \href{mailto:hasan.oguz@okan.edu.tr}{hasan.oguz@okan.edu.tr}; ORCID: \href{https://orcid.org/0000-0001-7484-4415}{0000-0001-7484-4415}}
\date{}

\begin{document}
\maketitle

\begin{abstract}
Third-order dispersion turns the soliton of the nonlinear Schr\"odinger equation, which underlies optical analogues of event horizons in fibers, into a weakly radiating solitary wave. We compute three of its properties without assuming a thermal emission law. First, the linearized (Bogoliubov--de Gennes) problem must be posed about the stationary state of the dispersive equation: linearizing about the integrable profile produces a complex quartet that is converged in grid and box, grows as the square root of the dispersion strength and is spurious. About the stationary state, on a periodic grid, the zero-mode sector consists of two Jordan blocks of size two and the spectrum is real at the operating point, including negative-energy box modes of the resonant branch. Second, we measure the steady Cherenkov emission of adiabatically prepared solitons from the flux of the radiation behind them, over eight decades. The radiation is phase-matched in the frame of the drifting soliton, and the first-order rate falls short of the measured one by factors of 29 to 45. Expressed through a scaled dispersion coefficient of the local soliton, the measured rate describes launched solitons of three amplitudes; with norm and momentum balance, and no further fitted parameter, it reproduces their recoil and the slow decay of their loss rate; at weak dispersion the loss of a launched soliton is the shedding of its launch mismatch. Third, a census of the asymptotic channels shows that long-wavelength upstream emission requires the chemical potential of the soliton, lowered by the comoving-frame Doppler shift of a drifting soliton, to lie inside the analysis band, which bounds the soliton amplitude. In a kinematic model of the flow the surface gravity has a closed form; the thermal relation and the entanglement built on it are imposed inputs of that model. The construction applies to Raman-free Kerr media.
\end{abstract}

\section{Introduction}
\label{sec:intro}

In an analogue event horizon, a transcritical flow reproduces the kinematics of black-hole radiance \cite{Hawking1974}: where the flow speed crosses the local excitation speed, positive- and negative-norm modes mix, and the mixing is thermal at a temperature set by the surface gravity of the horizon \cite{Unruh1981,Visser1998,Barcelo2011}. The prediction survives the trans-Planckian problem of the derivation: short-distance dispersion modifies the mode trajectories but not, to leading order, the thermal spectrum \cite{Jacobson1991,Unruh1995,Brout1995,Coutant2012,Macher2009,Robertson2012}, provided the dispersion is monotonic on the scale set by the surface gravity; a non-monotonic (roton--maxon) dispersion opens additional negative-energy branches that render the spectrum strongly nonthermal and permit emission without a horizon \cite{HolandaRibeiro2023}. Experimentally, moving refractive-index fronts in optical fibers and bulk media provided the first optical horizons \cite{Philbin2008,Belgiorno2010,Rubino2012,Drori2019}, following proposals based on slow light \cite{Leonhardt2000}; stimulated emission was measured in water waves \cite{Weinfurtner2011}, and experiments in atomic Bose--Einstein condensates, proposed as analogues in reference~\cite{Garay2000}, observed spontaneous emission with a thermal spectrum, entangled Hawking pairs and stationary emission \cite{Steinhauer2016,deNova2019,Kolobov2021}. In fibers, analogue Hawking radiation stimulated by a single photon has since been measured \cite{FelipeElizarraras2026}, and the backreaction of stimulated Hawking radiation has been studied in an optical analogue \cite{Procopio2026}; reference~\cite{Kranas2025} introduces nonlinear fiber optics and its gravitational analogues. The same Bogoliubov mechanism also operates without a horizon, in explicitly time-dependent analogue backgrounds \cite{Fischer2004_Cosmo}. For dispersive index fronts the emitted spectrum is not fixed by the surface gravity alone; the profile of the front and the dispersion relation enter the mode conversion \cite{Linder2016,Jacquet2015,Bermudez2016}. A construction that fixes the spectrum by a kinematic $\kappa$ is therefore a model assumption, and we treat it as one.

For solitonic horizons in dispersive optics, a complementary thermodynamic description was recently proposed: the spectral partition of the field into a coherent solitonic subsystem and an incoherent radiative bath defines operational entropies whose sum obeys a coarse-grained generalized second law (GSL) \cite{Bekenstein1973,Bekenstein1974,Bekenstein1981,Wall2010}, with integrability-breaking resonant radiation proposed there as the entropy carrier \cite{Oguz2026}. That analysis was restricted to undriven configurations, where the horizon inherits its structure from a freely evolving higher-order soliton. A pump that sustains such a horizon against radiative decay, and slow light for the probe that meets it, are natural steps toward a persistent laboratory horizon. They also sharpen the standard objection to analogue thermality: if the source is pumped, a thermal-looking fit to the emitted spectrum may simply re-encode the pump \cite{Barcelo2011,Weinfurtner2011}. Horizons in which pump and loss are explicit in the dynamics exist in driven-dissipative fluids of light, where the Hawking correlations are computed from the driven equations themselves \cite{Nguyen2015,Grisins2016}. For solitonic horizons no comparable driven treatment exists, and this paper does not provide one.

The soliton itself is a classic object of fiber optics. Third-order dispersion phase-matches the fundamental soliton to a dispersive wave, into which it radiates \cite{Wai1990,Karpman1993,Akhmediev1995,Dudley2006,Skryabin2010}: the soliton becomes a weakly nonlocal solitary wave, whose resonant tail lies beyond all orders of an expansion in the dispersion strength \cite{Boyd1998,Karpman2001}, while third-order equations with additional nonlinear dispersive terms also admit embedded solitons that do not radiate \cite{Champneys2001,Pelinovsky2005}. The recoil of the radiation shifts the spectrum of the soliton and can compensate its Raman self-frequency shift \cite{Skryabin2003,Biancalana2004}. The interaction of a probe with a soliton at a fiber event horizon can be described as cascaded four-wave mixing \cite{Webb2014}; a soliton can form such a horizon with its own dispersive wave \cite{Wang2015}, Raman deceleration of the soliton shifts the frequencies produced at the horizon \cite{Robertson2010}, and in a nonlinear waveguide four-wave mixing between a soliton and a continuous wave enhances the analogue Hawking pair production by orders of magnitude \cite{Robertson2019}. Against this background, the results below on the linearized problem, on the steady emission and the recoil of the soliton, and on the channels of long-wavelength emission concern the soliton that supports the horizon; the thermal relation enters only as an input of the kinematic model.

We compute three properties of the soliton that supports such a horizon; none of them relies on the thermal relation. First, the linearized problem has to be posed about the correct background. Third-order dispersion deforms the soliton at $\mathcal{O}(\dtres)$, so the integrable profile solves the equation only to that order, and its residual splits the zero-mode sector at $\mathcal{O}(\sqrt{\dtres})$ into a complex quartet, $|\mathrm{Im}\,\Omega|\simeq0.68N^{5/2}\sqrt{\dtres}$, that is converged in grid and box and yet spurious. About the stationary TOD state on the periodic computational domain, a Krein-signed Bogoliubov--de Gennes (BdG) analysis finds the zero-mode sector at the origin as two $2\times2$ Jordan blocks and a real spectrum at the operating point, including the negative-energy modes of the resonant branch, which are box modes (section~\ref{sec:bdg}). Second, we measure the steady Cherenkov emission of the soliton. A soliton prepared adiabatically settles to a steady emission rate, which we obtain from the flux of the resonant radiation that it leaves behind, over eight decades of the rate; the radiation is phase-matched in the frame of the drifting soliton, and the first-order rate falls short of the measured one by factors of $29$ to $45$. Expressed through the scaled dispersion coefficient of the local soliton, the measured rate also describes launched solitons; with norm and momentum balance, and no further fitted parameter, it accounts for their recoil and for the slow algebraic decay of their loss rate, $-d\ln Q/d\xi\simeq c/\xi$ for the core norm $Q$ (section~\ref{sec:emission}). At weak dispersion the loss of a launched soliton is instead the shedding of its launch mismatch (section~\ref{sec:bdg}). Third, a census of the asymptotic channels shows at which frequencies a long-wavelength Hawking mode could escape upstream. For a soliton at rest this requires $\omega>\mu=N^2/2$, and hence $N<\sqrt{2\omega_{\rm hi}}$ within an analysis band that ends at $\omega_{\rm hi}$; in the frame comoving with a soliton drifting at $V_d$ the threshold is lowered by $V_d^2/2$. The limit is set by the chemical potential of the soliton, not by the flow. At the operating point the kinematic Hawking window, the part of the band in which the Hawking mode can escape upstream and its negative-norm partner can propagate downstream, covers $71\%$ of the width of the analysis band but only $3\%$ of the pair flux that the imposed thermal relation assigns to the band (section~\ref{sec:thermality}).

The analogue-horizon layer is a kinematic model of the flow, with a phenomenological two-band slow-light factor and a drive-translated horizon (section~\ref{sec:model}), in which the surface gravity $\kappa$ is defined at the group-velocity crossing and never from an emission diagnostic. The first two results do not use this model, and the third uses only its downstream flow speed. Four statements then hold by construction and are properties of the model, not findings: $\kappa$ does not depend on the drive, which enters only through the horizon position; $\kappa=N\,S(k_{\rm op})$ in closed form, so its dependence on the slow-light gap is that of $S$; a log-linear fit of the near-horizon thermal ratio returns the $\kappa$ that generated it; and the logarithmic negativity $E_N=2\,\operatorname{arcsinh}|\beta_\omega|$ follows from that assumed ratio. The numerical evaluation adds an error budget to the first statement: across a $\pm30\%$ drive range the extracted $\kappa$ varies by $1.6\times10^{-5}$ in relative terms, inside a closed-form discretization band of width $1.8\times10^{-5}$ that falls as $\Delta\tau^{2}$ (section~\ref{sec:kappa}). The thermal relation and the entanglement built on it remain model inputs: a scattering calculation for an operator assembled from the soliton and the model flow does not reproduce the relation (section~\ref{sec:discussion}). The kinematic construction is therefore not a derivation of Hawking thermality; it is a controlled framework in which the consequences of an independently specified surface gravity can be compared with the channel structure of the soliton background.

We also evaluate the Gaussian covariance of the idealized Hawking pairs of the model, which a persistent horizon would make experimentally relevant. Each pair with the imposed Bogoliubov amplitudes $(\alpha_\omega,\beta_\omega)$, acting on vacuum, produces a two-mode squeezed state whose partial transpose is negative at every frequency, with logarithmic negativity $E_N=2\,\operatorname{arcsinh}|\beta_\omega|$ \cite{Vidal2002,Plenio2005,Simon2000,Adesso2007}. Nonseparability of the Hawking mode and its partner is the observable that a condensate experiment measured \cite{Steinhauer2016} and that theory identifies as the signature of spontaneous pair production \cite{Finazzi2014,Busch2014}; for phonon pairs in a quantum fluid of light it has been analyzed together with its detection through correlations of the emitted light \cite{BuschCarusotto2014}. Within the model only the magnitude of the negativity carries information, because its sign is fixed; we evaluate it for the idealized two-mode pair at an independently fixed $\kappa$, as a diagnostic of the model and not as a dynamical prediction (section~\ref{sec:entanglement}).

\section{Kinematic model and linearization}
\label{sec:model}

We work with the dimensionless GNLSE of the anomalous-dispersion regime, with $\xi=z/L_D$, $\tau=-T/T_0$, $\psi=(\gamma L_D)^{1/2}A$, $L_D=T_0^2/|\beta_2|$ and $\dtres=\beta_3/(6|\beta_2|T_0)$ \cite{Agrawal2013},
\begin{equation}
\partial_\xi \psi \;=\; i\!\left(\tfrac12\,\partial_\tau^2 \psi + |\psi|^2\psi + i\,\dtres\,\partial_\tau^3\psi\right),
\label{eq:gnlse}
\end{equation}
where $\xi$ is propagation distance in dispersion lengths, $\tau$ the comoving time and $\dtres$ the third-order-dispersion (TOD) strength; the TOD term enters with the sign convention $+i\dtres\partial_\tau^3$ of reference~\cite{Oguz2026} \cite{Agrawal2013,Akhmediev1995,Skryabin2010}. The undriven fundamental background is $\psi_0(\tau)=N\operatorname{sech}(N\tau)$ with chemical potential $\mu=N^2/2$; in physical units it has width $T_0/N$ and peak power $P_0=N^2|\beta_2|/(\gamma T_0^2)$, and a wavenumber $k$ in $\tau$ is a frequency offset $k/T_0$, positive on the blue side. Linear waves obey $\beta_0(k)=k^2/2-\dtres k^3$ with bare group velocity $v_{g0}(k)=k-3\dtres k^2$.

The kinematic model adds two ingredients to equation~\eqref{eq:gnlse}: a two-band slow-light factor and a drive-translated horizon. (i) A phenomenological slow-light factor with coupling gap $\kg$, motivated by the gap of a Bragg resonance \cite{Mok2006} and modeled at the two-band level by the suppression factor
\begin{equation}
S(k;\dtres,\kg)\;=\;\frac{|\beta_0(k)|}{\sqrt{\beta_0(k)^2+\kg^2}},
\qquad S\to 1 \text{ as } \kg\to0,
\label{eq:slowlight}
\end{equation}
which renormalizes the group velocity to $v_g=v_{g0}\,S$. The factor interpolates smoothly between strong suppression near the band edge, $S\to|\beta_0|/\kg$ as $\beta_0\to0$, and transparency far from it, $S\to1$. It is a two-band model, not a derived dispersion relation; its microscopic counterpart is the dispersion-computed $v_g$ of the full linearized problem (section~\ref{sec:discussion}). Since $S$ vanishes wherever $\beta_0=0$, at the carrier and near the resonant wavenumber $1/(2\dtres)$, it cannot be read as a grating acting on the spectrum of the soliton: it is a phenomenological factor applied to the flow at $k_{\rm op}$, where $S=0.954$ at the operating point below. (ii) A drive of normalized amplitude $d$ (pump power relative to the soliton-sustaining value $d=1$) that maintains the localized background against radiative decay.

The effective flow seen by counterpropagating fluctuations is modeled by the kinematic ansatz
\begin{equation}
v(\tau)\;=\;\cmatch\!\left[\,1+\tanh\!\frac{\kappa_0(\tau-\tau_h)}{\cmatch}\,\right],
\label{eq:flow}
\end{equation}
with $\kappa_0=N\,S(k_{\rm op})$, matching speed $\cmatch=\max\{|v_{g0}(k_{\rm op})|\,S(k_{\rm op}),\,10^{-3}\}$ at the operating wavenumber $k_{\rm op}$, and horizon position
\begin{equation}
\tau_h=-\frac{d-1}{2N}.
\label{eq:tauh}
\end{equation}
Equation~\eqref{eq:flow} is an explicit model of the flow: the slow-light factor enters only through the rescaling of $\kappa_0$ and $\cmatch$, and the drive enters only through the horizon position, equation~\eqref{eq:tauh}. No pump term is added to equation~\eqref{eq:gnlse}, and the linear response of section~\ref{sec:bdg} is that of the undriven soliton. The gradient scale $\kappa_0=N\,S(k_{\rm op})$ is not a free parameter. The background $\psi_0=N\operatorname{sech}(N\tau)$ has inverse width $N$, so $N$ is the scale the soliton itself sets, and $S$ is the slow-light rescaling of that scale. Equation~\eqref{eq:kappa} below recovers $\kappa_0$ from the profile instead of defining it independently, and the closed form of section~\ref{sec:kappa} makes the recovery exact in the continuum. The linear translation law~\eqref{eq:tauh} is a hypothesis of the model and is not implied by the linearization: a spatially symmetric pump has no overlap with the odd translational mode $\psi_0'$ and couples instead to the even width deformation $\partial_N\psi_0$, which would change the flow gradient and lies in the zero-frequency generalized kernel rather than above the gap $\mu$. Equation~\eqref{eq:tauh} therefore presupposes a drive that translates the soliton while holding its amplitude fixed, which a symmetric pump does not do at linear order. Section~\ref{sec:kappa} evaluates its consequence, a drive-independent $\kappa$, which within the model is an identity; testing the hypothesis itself requires a driven model and the dispersion-computed group velocity (section~\ref{sec:discussion}).

The horizon is the group-velocity crossing $v(\tau_h^{\times})=\cmatch$, and the surface gravity is defined kinematically,
\begin{equation}
\kappa \;\equiv\; \left|\frac{dv}{d\tau}\right|_{\tau=\tau_h^{\times}},
\label{eq:kappa}
\end{equation}
evaluated by second-order finite differencing of $v(\tau)$ on a uniform grid of $4096$ points over $\tau\in[-20,20]$ with linear interpolation to the crossing. No emission observable, signal amplitude or fitted quantity enters equation~\eqref{eq:kappa}. The associated temperature $T_H=\kappa/2\pi$ is the \emph{local kinematic surface-gravity temperature} of the crossing, a property of the flow gradient, and is not an emission temperature of the device.

The operating point is
\begin{equation}
N=1,\qquad \dtres=0.05,\qquad \kg=0.3,\qquad k_{\rm op}=1.5,
\label{eq:oppoint}
\end{equation}
and all results refer to it unless stated otherwise. There the chemical potential $\mu=\tfrac12$ lies inside the analysis band $\omega\in[0.05,1.6]$, so that emission to the upstream side is kinematically allowed, and the flow is fast enough for the pair channel, in which the negative-norm partner propagates downstream, to cover the band (section~\ref{sec:thermality}). The value $N\dtres=0.05$ lies inside the range in which the stationary TOD state is localized on the BdG grid; section~\ref{sec:bdg} shows its spectrum. The radiative loss and the channel census are evaluated around the operating point, over $N\dtres\in[0.012,0.10]$ and over the $(N,k_{\rm op})$ plane, and the steady emission and the recoil over $N\dtres\in[0.035,0.12]$ (section~\ref{sec:emission}).

\section{Results}
\label{sec:results}

\subsection{BdG spectrum and radiative loss}
\label{sec:bdg}

\begin{figure}[tbp]
\centering\includegraphics{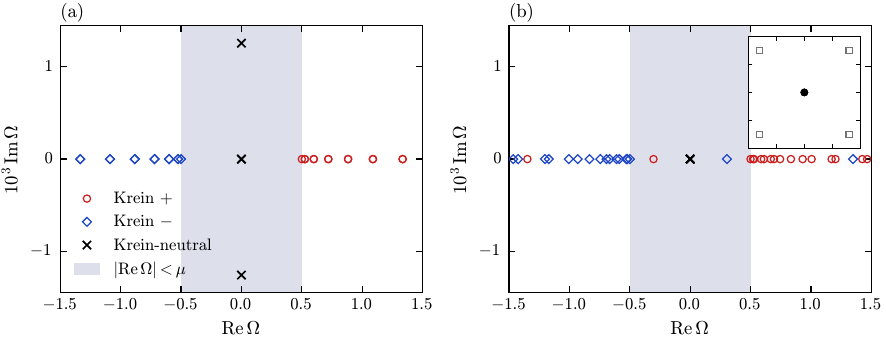}
\caption{Krein-signed BdG spectra of the $N=1$ soliton ($\mu=0.5$; $n_\tau=500$, $\tau_{\max}=16$) in the window $|\mathrm{Re}\,\Omega|\le3\mu$. (a)~$\dtres=0$, linearization about $\psi_0$. (b)~The operating value $\dtres=0.05$, linearization about the stationary periodic-box TOD state $\phi$; inset ($|\mathrm{Re}\,\Omega|,|\mathrm{Im}\,\Omega|\le0.2$, ticks at $0$ and $\pm0.1$): the zero-mode sector about $\phi$ (filled circle) and the complex quartet about $\psi_0$ at the same $\dtres$ (open squares). Main panels: circles, positive symplectic norm (Krein-positive); diamonds, negative norm (Krein-negative); crosses, Krein-neutral eigenvalues (the zero-mode sector and every non-real eigenvalue); shaded band: $|\mathrm{Re}\,\Omega|<\mu$.}
\label{fig:krein}
\end{figure}

Linearizing equation~\eqref{eq:gnlse} about $\psi_0=N\operatorname{sech}(N\tau)$ with $\psi=e^{i\mu\xi}[\psi_0+u\,e^{-i\Omega\xi}+v^*e^{i\Omega^*\xi}]$ gives the BdG eigenproblem $\Omega\,(u,v)^T=M\,(u,v)^T$ with
\begin{equation}
M=\begin{pmatrix} H-i\dtres\partial_\tau^3 & -\psi_0^2\\[2pt] \psi_0^2 & -H-i\dtres\partial_\tau^3\end{pmatrix},
\quad
H=-\tfrac12\partial_\tau^2-2\psi_0^2+\mu.
\label{eq:bdg}
\end{equation}
The TOD term acts identically on both components, because it originates as a linear-dispersion correction. Each eigenvector carries a symplectic norm of sign $\sigma=\operatorname{sign}(\|u\|^2-\|v\|^2)$, its Krein signature, and its energy has the sign of $\sigma\,\mathrm{Re}\,\Omega$. In a Hamiltonian BdG problem an instability can arise through the collision of real eigenvalues of opposite energy sign, through the generalized kernel at the origin, or from negative-energy eigenvalues embedded in the continuum, and, when the Hessian of the energy has finitely many negative directions, the Hamiltonian--Krein index bounds the number of eigenvalues that can take part \cite{Kapitula2004}; here the grid modes of the resonant branch with $k>k_{\rm RR}$ have negative energy as well, so the signature classifies the computed real eigenvalues and does not bound their number. The census of norms and energies, together with the rank structure of the generalized kernel, is therefore the natural stability diagnostic for the soliton that supports the horizon; the kinematic flow is not part of this operator. Because $\sigma_3M$ is Hermitian, with $\sigma_3$ the third Pauli matrix acting on $(u,v)$, an eigenvector with $\mathrm{Im}\,\Omega\neq0$ has $\|u\|^2=\|v\|^2$: non-real eigenvalues are Krein-neutral, and the signature classifies the real spectrum only.

Figure~\ref{fig:krein} shows the two regimes. At $\dtres=0$ [figure~\ref{fig:krein}(a)] the problem is the linearization about the integrable NLSE soliton, whose spectrum is entirely real, so any nonzero imaginary part at $\dtres=0$ is discretization error. The computed spectrum is real to machine precision on the diagnostic annulus $0.1\mu<|\mathrm{Re}\,\Omega|\le3\mu$, where $\max|\mathrm{Im}\,\Omega|<10^{-15}$ and the continuum edge sits at the gap $\mu$ with relative error $2.3\times10^{-7}$. The inner cutoff of the annulus removes the lattice images of the zero-mode sector. Because $\psi_0$ solves the discretized equation only to $\mathcal{O}(h^4)$, with $h$ the grid spacing, the four eigenvalues of the sector split at $\mathcal{O}(h^2)$ into a real doublet at $\mathrm{Re}\,\Omega\simeq\pm5.2\times10^{-4}$ and a spurious purely imaginary pair at $|\mathrm{Im}\,\Omega|\simeq1.3\times10^{-3}$; none of the four carries Krein imbalance (they are drawn as Krein-neutral in figure~\ref{fig:krein}(a)), and all four collapse under refinement (supplementary material, section S4). The bound above therefore refers to the physical sector, not to every eigenvalue of the discretized operator. On the annulus the symplectic norms are balanced, $13$ positive against $13$ negative, as the symmetry $\sigma_1M^*\sigma_1=-M$ requires, and every eigenvalue carries positive energy. The reality of the annulus spectrum follows from the structure of the discretized operator, since $\Omega^2\in\mathrm{spec}(L_-L_+)$ with $L_\pm=-\tfrac12\partial_\tau^2+\mu-(2\pm1)\psi_0^2$, and it persists on grids too coarse to resolve the soliton. The quantitative check of the discretization is the reproduction of the gap.

Switching on TOD [figure~\ref{fig:krein}(b)] deforms the background itself. The profile $\psi_0$ no longer solves equation~\eqref{eq:gnlse}, since $i\dtres\partial_\tau^3\psi_0\neq0$, and the linearization must be taken about the stationary TOD state $\phi$ on the periodic computational domain, defined by $\tfrac12\phi''+|\phi|^2\phi+i\dtres\phi'''=\mu\phi$ and obtained by Newton iteration on the grid of figure~\ref{fig:krein}. On this periodic grid $\phi$ carries a standing resonant tail, $3.6\times10^{-5}$ of its amplitude at the operating point, and a mean wavenumber $\dtres N^2$ at leading order: it is the member of the soliton family at rest in $\tau$, whereas the launched soliton considered in the loss simulations below is the member of zero mean wavenumber, which drifts (supplementary material, section S5). About $\phi$ the operator of equation~\eqref{eq:bdg} carries $-\phi^2$ and $\phi^{*2}$ in its off-diagonal blocks and $H=-\tfrac12\partial_\tau^2-2|\phi|^2+\mu$; the zero-mode sector stays at the origin with the structure it has in the integrable limit, two $2\times2$ Jordan blocks formed by the phase and translation modes and their generalized partners. The singular values of the discretized $M$ and $M^2$ give $\dim\ker M=2$ and $\dim\ker M^2=4$, the null singular values lying at least eight orders of magnitude below the next, and random perturbations of $M$ split the sector as the square root of their size, which fixes both blocks at size two (supplementary material, section S4). The four computed eigenvalues of the sector lie below $10^{-6}$ [the filled circle in the inset of figure~\ref{fig:krein}(b)], at the $\mathcal{O}(\sqrt{\epsilon_{\rm mach}\rho})$ level that a backward-stable eigensolver returns for a defective eigenvalue ($\epsilon_{\rm mach}$ the machine precision, $\rho$ the spectral radius of $M$). At the operating value $\dtres=0.05$ the grid resolves the resonant branch, whose far-field dispersion $\tfrac12k^2+\mu-\dtres k^3$ changes sign near $k_{\rm RR}\simeq10$ (at $\dtres=0.02$ the corresponding $k_{\rm RR}\simeq25$ lies beyond what the difference stencils represent on this grid). The annulus then holds $34$ eigenvalues, $17$ of each norm, all real to $|\mathrm{Im}\,\Omega|<10^{-13}$. Four of them carry negative energy: the positive-norm modes of the resonant branch at $\mathrm{Re}\,\Omega=-0.30$ and $-1.35$ and their negative-norm mirror images at $+0.30$ and $+1.35$, the two circles at $\mathrm{Re}\,\Omega<0$ and the two diamonds at $\mathrm{Re}\,\Omega>0$ in figure~\ref{fig:krein}(b). These are the eigenvalues through which a collision of opposite energy signs could destabilize the soliton, and at the operating point none has occurred on this grid. Since they are box modes, their positions, and with them the absence of collisions, refer to the box $\tau_{\max}=16$, as does their number. On the annulus and in the box $\tau_{\max}=16$, complex eigenvalues appear only for $N\dtres\ge0.08$ (the onset depends on the box), outside the range $N\dtres\le0.05$ in which the BdG spectra are evaluated, where the standing tail of the periodic stationary state has grown to between $0.5\%$ and $9\%$ of its amplitude (supplementary material, section S4). Linearizing instead about $\psi_0$ splits the zero-mode sector into a complex quartet with $|\mathrm{Im}\,\Omega|\simeq0.68N^{5/2}\sqrt{\dtres}$ [inset of figure~\ref{fig:krein}(b)]: the $\mathcal{O}(\dtres)$ residual of $\psi_0$ splits the two Jordan blocks at $\mathcal{O}(\sqrt{\dtres})$, and the power of $N$ follows from the exact amplitude scaling. The quartet is grid- and box-converged but absent about $\phi$; it is an artifact of the non-stationary background and is excluded from the stability analysis. It is the generic square-root splitting of a $2\times2$ Jordan block by a perturbation of order $\dtres$: a caution against linearizing about the integrable profile, not an instability. On the infinite line the zero eigenvalue of the TOD soliton is embedded in the continuous spectrum, since the branch $\tfrac12k^2+\mu-\dtres k^3$ vanishes near $k\simeq10.1$ and decreases without bound beyond it: there the soliton radiates (section~\ref{sec:emission}), the double eigenvalue becomes a resonance whose width is set by the emission, and the Jordan blocks and the real spectrum found here are properties of the periodic problem in the box $\tau_{\max}=16$.

The physical effect of TOD is the resonant (Cherenkov) channel at the wavenumber $k_{\rm RR}$ phase-matched to the soliton \cite{Wai1990,Akhmediev1995,Skryabin2010,Dudley2006,Husko2013}. For the soliton of zero mean wavenumber, which drifts at $-\dtres N^2$ and is the one launched below, the condition is $\dtres k_{\rm RR}^3-\tfrac12k_{\rm RR}^2-\dtres N^2k_{\rm RR}=\tfrac12N^2$, which gives $k_{\rm RR}=10.19$ at the operating point; for the soliton at rest the drift term is absent and $k_{\rm RR}=10.10$ (section~\ref{sec:emission}). On a periodic grid this channel appears only through box-dependent collisions with discretized continuum modes, so the loss is measured by direct integration of equation~\eqref{eq:gnlse} with absorbing boundaries, launched from $\psi_0$ and followed to $\xi=40$ (supplementary material, section S5), which is $6.1$ surface-gravity periods $2\pi/\kappa$ at the operating point, where $\kappa=0.954$ (section~\ref{sec:kappa}). Table~\ref{tab:loss} lists, for each value of $N\dtres$ considered, the fraction of the norm $Q(\xi)$ within $6/N$ of the intensity maximum that is lost over $\xi\in[0,40]$, and the drift of the center $\tau_c$, the sub-grid intensity maximum, over $\xi\in[30,40]$. Two regimes appear, and in neither does the soliton radiate at a constant rate.

\begin{table}[tbp]
\centering
\caption{Radiative loss of the launched soliton $\psi_0=N\operatorname{sech}(N\tau)$ under equation~\eqref{eq:gnlse}: fraction of the core norm lost over $\xi\in[0,40]$ and drift of the center over $\xi\in[30,40]$. The operating point~\eqref{eq:oppoint} is the entry $N=1$, $N\dtres=0.05$.}
\label{tab:loss}
\begin{tabular}{cclr}
\toprule
$N\dtres$ & $N$ & $1-Q(40)/Q(0)$ & $d\tau_c/d\xi$\\
\midrule
0.02 & 1 & $1.4\times10^{-4}$ & $-0.020$\\
0.03 & 1 & $3.1\times10^{-4}$ & $-0.030$\\
0.04 & 1 & $5.7\times10^{-4}$ & $-0.041$\\
0.05 & 1 & $9.3\times10^{-4}$ & $-0.052$\\
0.05 & 2 & $9.4\times10^{-4}$ & $-0.104$\\
0.06 & 1 & $1.4\times10^{-3}$ & $-0.064$\\
0.08 & 1 & $5.5\times10^{-3}$ & $-0.103$\\
0.10 & 1 & $2.3\times10^{-2}$ & $-0.200$\\
0.10 & 2 & $3.5\times10^{-2}$ & $-0.526$\\
\bottomrule
\end{tabular}
\end{table}

\begin{figure}[tbp]
\centering\includegraphics{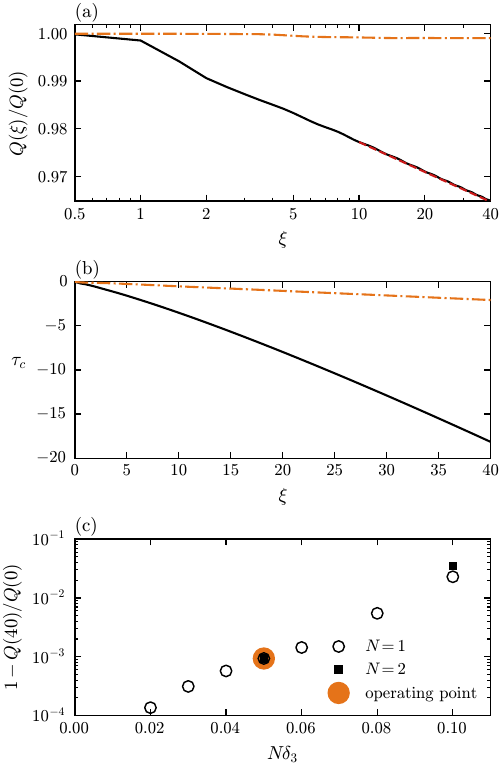}
\caption{Radiative loss of the launched soliton. (a)~Core norm $Q(\xi)/Q(0)$ on logarithmic axes for $N=2$, $\dtres=0.05$ ($N\dtres=0.10$; solid) and at the operating point $N=1$, $\dtres=0.05$ (dash-dotted), with the law $Q\propto\xi^{-c}$ through $Q(10)$ for the former (dashed) with $c=9.2\times10^{-3}$. (b)~Center $\tau_c(\xi)$, the sub-grid intensity maximum, for the same two simulations. (c)~Norm lost over $\xi\in[0,40]$ against $N\dtres$ for $N=1$ (open circles) and $N=2$ (filled squares); large disk: the operating point.}
\label{fig:loss}
\end{figure}

For $N\dtres\le0.06$ the lost norm is $(0.33$--$0.40)(N\dtres)^2$ at each of the sixteen points computed in this range, and the drift follows, to within $7\%$, the leading-order value $-\dtres N^2$ of the TOD soliton of zero mean wavenumber (supplementary material, section S5). The launched profile $\psi_0$ differs from that soliton at $\mathcal{O}(N\dtres)$, and it sheds the mismatch as radiation during the launch transient. The steady emission of the soliton is smaller by orders of magnitude (section~\ref{sec:emission}); the standing tail of $3.6\times10^{-5}$ of the periodic stationary state is not a measure of it, since at fixed grid spacing the tail depends on the length of the box. Beyond $N\dtres\simeq0.06$ the loss outgrows the quadratic law, reaching $0.85(N\dtres)^2$ at $N\dtres=0.08$ and $2.3(N\dtres)^2$ at $0.10$ for $N=1$, and the drift exceeds $\dtres N^2$ in magnitude. There the emitted radiation carries momentum away, and the recoil moves the mean wavenumber of the soliton away from phase matching \cite{Skryabin2010,Skryabin2003,Biancalana2004}, so that the loss rate decays with time while the drift grows in magnitude. At $N\dtres=0.10$ the decay is algebraic at late times, $-d\ln Q/d\xi\simeq c/\xi$. The two amplitudes of table~\ref{tab:loss} at $N\dtres=0.10$ are related by the exact scaling $\psi(\tau,\xi)=N\Psi(N\tau,N^2\xi)$ with $\tilde\dtres=N\dtres$, under which the $N=2$ simulation at $\dtres=0.05$ is equivalent to the $N=1$ simulation at $\dtres=0.10$ extended to $\xi=160$; the $N=2$ simulation yields $c=9.2\times10^{-3}$ from the windows $\xi\in[10,20]$, $[20,30]$ and $[30,40]$ [the dashed line of figure~\ref{fig:loss}(a)], and at $N\dtres=0.08$ the corresponding pair, $(N,\dtres)=(1,0.08)$ and $(1.6,0.05)$, yields $c=1.9\times10^{-3}$ and $3.1\times10^{-3}$, so that there the exponent continues to evolve over the simulated interval (supplementary material, section S5). Section~\ref{sec:emission} derives this decay and these values of $c$ from the measured emission rate and the balance of norm and momentum. The $N=2$ simulation at $N\dtres=0.10$ shows the recoil-limited regime most clearly (figure~\ref{fig:loss}): the core loses $3.5\%$ of its norm over $\xi\le40$, and the drift, the local slope of $\tau_c(\xi)$ in figure~\ref{fig:loss}(b), grows in magnitude from $-0.31$ to $-0.53$ in the frame of the carrier group velocity.

At the operating point, $N\dtres=0.05$, the loss belongs to the first regime and is confined to the launch transient. Of the $9.3\times10^{-4}$ lost over $\xi\le40$, $7.8\times10^{-4}$ is lost by $\xi=10$ and $1.3\times10^{-4}$ over the surface-gravity period that follows, still the shedding of the launch mismatch since the prepared soliton radiates only $3\times10^{-8}$ of its norm per surface-gravity period (section~\ref{sec:emission}), after which $Q$ is constant to within an oscillation of $5\times10^{-5}$ peak to peak; the entry $N=2$, $N\dtres=0.05$ of table~\ref{tab:loss}, which by the same scaling is the operating-point simulation extended to $\xi=160$, has lost $9.4\times10^{-4}$. The drift settles from $-0.042$ in $\xi\in[0,5]$ to $-0.052$ in $\xi\in[30,40]$, within $4\%$ of $-\dtres N^2$. The background is therefore quasi-static in norm and velocity on the timescale of the kinematic construction; the channel analysis of section~\ref{sec:thermality} includes this drift. The far field is spectrally stable, $\Omega_\pm(k)=\pm(\tfrac12 k^2+\mu)-\dtres k^3\in\mathbb{R}$. This does not exclude a black-hole-laser instability, which is a global mode of a black-hole/white-hole pair \cite{CorleyJacobson1999,Faccio2012}: the ansatz~\eqref{eq:flow} has a single crossing and no such cavity, while the profile of a physical soliton has two, and their pair is not analyzed here. In soliton telecommunications third-order dispersion is a parasitic distortion to be engineered away; here it opens the only radiative channel. The resonant branch thus plays the role of the rotonic branch of dipolar condensates \cite{HolandaRibeiro2023}: a negative-energy branch created by the short-distance dispersion that radiates whenever a conversion mechanism is present, with or without a horizon; here the conversion is the classical phase matching to the soliton.

\subsection{Steady Cherenkov emission and recoil}
\label{sec:emission}

The launched profile $\psi_0$ is not a soliton of equation~\eqref{eq:gnlse}, so the loss of table~\ref{tab:loss} mixes the shedding of the launch mismatch with the emission of the soliton itself. To measure the emission we prepare the soliton adiabatically: $\psi_0$ is launched at $\dtres=0$, $\dtres$ rises as $\dtres[1-\cos(\pi\xi/100)]/2$ to its final value at $\xi=100$ and is then held, with the same numerical scheme, grid and absorber as the loss simulations (supplementary material, section S5). At the weaker dispersions the rate lies far below what the core norm resolves, but the radiation that the soliton leaves behind is a plane wave of uniform intensity $I$ at a single wavenumber $k_b$, and the emitted norm flux $I\,|v_{g0}(k_b)-V|$, with $V$ the velocity of the soliton, divided by $Q$ is the relative emission rate $R=-d\ln Q/d\xi$. Table~\ref{tab:steady} lists $R$ for $N=1$. Over the hold phase the emission is steady: $R$ is constant to $6\times10^{-5}$ in relative terms at $\dtres=0.035$ and to $2\%$ at $\dtres=0.07$, and it spans eight decades, from $2.8\times10^{-14}$ to $8.2\times10^{-6}$ per unit $\xi$. Where the rate exceeds $10^{-9}$ (core norm) and $10^{-10}$ (box norm), the slopes of the core norm and of the norm left in the computational box agree with it to within $2\%$, and halving $\Delta\xi$, doubling the grid, enlarging the box together with its absorber, halving or doubling the ramp time, or filtering the spectrum changes it by at most $3.1\times10^{-4}$ in relative terms.

\begin{table}[tbp]
\centering
\caption{Steady emission of prepared solitons, $N=1$: emission rate $R$ from the flux of the radiation, as the mean over the hold phase $\xi\in[140,200]$, and its relative standard deviation there; wavenumber $k_b$ of the radiation; largest root $k_r=k_s+q_rN/\sqrt{D_2}$ of equation~\eqref{eq:resonance} for the measured amplitude and mean wavenumber, and the root of the condition without the drift term; ratio of $R$ to the first-order rate~\eqref{eq:born}.}
\label{tab:steady}
\footnotesize\setlength{\tabcolsep}{4pt}
\begin{tabular}{ccclllc}
\toprule
$\dtres$ & $R$ & spread & $k_b$ & $k_r$ & $k_r$, no drift & $R/R_{\rm B}$\\
\midrule
0.035 & $2.77\times10^{-14}$ & $6\times10^{-5}$ & $14.4251$ & $14.4237$ & $14.3550$ & $45.0$\\
0.040 & $4.52\times10^{-12}$ & $8\times10^{-5}$ & $12.6591$ & $12.6570$ & $12.5790$ & $42.2$\\
0.045 & $2.22\times10^{-10}$ & $9\times10^{-5}$ & $11.2900$ & $11.2869$ & $11.1997$ & $39.6$\\
0.050 & $4.76\times10^{-9}$ & $1\times10^{-4}$ & $10.1986$ & $10.1943$ & $10.0981$ & $37.2$\\
0.055 & $5.56\times10^{-8}$ & $2\times10^{-4}$ & $9.3092$ & $9.3035$ & $9.1984$ & $35.0$\\
0.060 & $4.14\times10^{-7}$ & $1\times10^{-3}$ & $8.5717$ & $8.5642$ & $8.4505$ & $32.8$\\
0.065 & $2.15\times10^{-6}$ & $6\times10^{-3}$ & $7.9521$ & $7.9424$ & $7.8205$ & $30.8$\\
0.070 & $8.19\times10^{-6}$ & $2\times10^{-2}$ & $7.4275$ & $7.4155$ & $7.2861$ & $29.0$\\
\bottomrule
\end{tabular}
\end{table}

The wavenumber of the radiation identifies the resonance. A soliton $a(\tau-V\xi)\,e^{ik_s\tau+i\theta\xi}$ of mean wavenumber $k_s$ drives the linear waves at the frequency $\theta-(k-k_s)V$, so that the radiation is phase-matched in the frame of the drifting soliton, $\beta_0(k)+\theta-(k-k_s)V=0$. About $k_s$ the dispersion is again quadratic and cubic, with curvature $D_2=1-6\dtres k_s$, and the exact scaling maps the local soliton of amplitude $N$ onto the zero-wavenumber soliton of unit amplitude at the dispersion coefficient $\tilde\delta=N\dtres/D_2^{3/2}$; with $\theta=-\beta_0(k_s)+N^2/2$ and the velocity $V=v_{g0}(k_s)-\dtres N^2/D_2$ of that soliton at leading order, the phase-matching condition reads
\begin{equation}
\tilde\delta q^3-\tfrac12q^2-\tilde\delta q-\tfrac12=0,\qquad q=\frac{(k-k_s)\sqrt{D_2}}{N},
\label{eq:resonance}
\end{equation}
in which the term $-\tilde\delta q$ is the drift; without it equation~\eqref{eq:resonance} is the condition of the soliton at rest. The measured $k_b$ agrees with the largest root $k_r$ to within $0.17\%$ at every $\dtres$ of table~\ref{tab:steady}, while the condition without the drift misses it by $0.5$ to $1.9\%$. To first order in $\dtres$ the source $-\dtres\psi_0'''$ radiates at the phase-matched wavenumber at the rate $|\hat S(k_r)|^2/|v_{\rm rel}|$, with $\hat S$ the Fourier transform of the source and $v_{\rm rel}$ the group velocity of the radiation relative to the soliton \cite{Karpman1993,Akhmediev1995}; relative to the norm,
\begin{equation}
R_{\rm B}=N^2\,\frac{\pi^2\tilde\delta^2q_r^6\operatorname{sech}^2(\pi q_r/2)}{2\,|q_r-3\tilde\delta q_r^2+\tilde\delta|}.
\label{eq:born}
\end{equation}
Its exponent $\pi q_r$, set by the poles of $\operatorname{sech}(N\tau)$ nearest the real axis, accounts for the variation of the measured rate: across table~\ref{tab:steady}, $\ln R$ changes by $19.5$ and $\ln R+\pi q_r$ by $2.4$. Its prefactor does not: the measured rate exceeds $R_{\rm B}$ by a factor that falls from $45$ at $\dtres=0.035$ to $29$ at $\dtres=0.07$. At the phase-matched wavenumber the third-order term of the dispersion is as large as the second-order one, so the first-order source does not fix the amplitude of the exponentially small radiation; it is a beyond-all-orders quantity \cite{Boyd1998,Karpman2001}, which the measurement supplies here.

\begin{figure}[tbp]
\centering\includegraphics{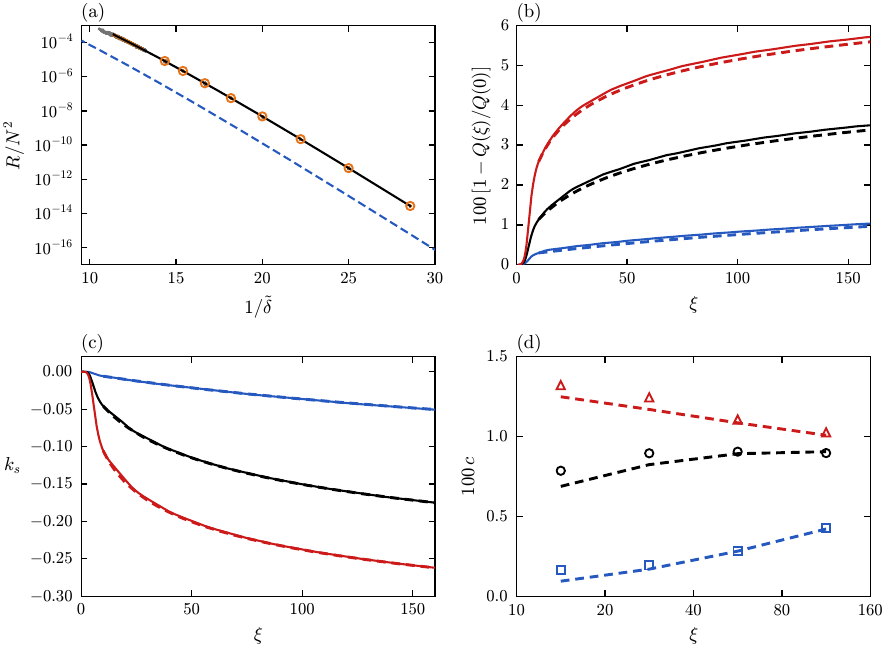}
\caption{Steady emission and recoil. (a)~Scaled emission rate $R/N^2$ against $1/\tilde\delta$: steady prepared solitons of table~\ref{tab:steady} (circles), every rate of the prepared solitons over their hold phase (orange dots; the data of the fit), the rates of the launched solitons at $N=1$, $1.6$ and $2$ for radiation emitted after $\xi=20$ (grey dots), the rate function~\eqref{eq:ratefn} (solid) and the first-order rate~\eqref{eq:born} (dashed). (b)~Norm lost by launched solitons at $N=1$ and $\dtres=0.08$ (blue), $0.10$ (black) and $0.12$ (red), direct integration (solid) and recoil model~\eqref{eq:recoil} started at $\xi=10$ (dashed). (c)~Mean wavenumber $k_s$ of the same simulations. (d)~Exponent $c$ of $-d\ln Q/d\xi\simeq c/\xi$ over the windows $[10,20]$, $[20,40]$, $[40,80]$ and $[80,160]$, at the geometric center of each window: simulations (symbols) and model (dashed).}
\label{fig:emission}
\end{figure}

If the emission is a property of the local soliton, $R/N^2$ depends on its state only through $\tilde\delta$. We represent this dependence by the empirical interpolation formula
\begin{equation}
R=N^2R_s(\tilde\delta),\qquad R_s(\tilde\delta)=C\,\tilde\delta^{\,p}\,e^{\,g\tilde\delta-\pi q_r(\tilde\delta)},
\label{eq:ratefn}
\end{equation}
fitted to every rate of the prepared solitons over their hold phase, for $\dtres$ from $0.035$ to $0.12$; above $\dtres=0.07$ the soliton loses enough norm during the ramp that its $\tilde\delta$ still decreases during the hold phase, so the rates of these simulations enter the fit point by point. The fit, with $C=151.3$, $p=-2.85$ and $g=-13.7$ (strongly correlated parameters without separate meaning), reproduces the rates to $0.7\%$ (rms) for $0.035\le\tilde\delta\le0.088$ [figure~\ref{fig:emission}(a)]. The launched solitons test this description, since they enter the dispersive equation with a mismatch, carry a breathing oscillation and, for $N\dtres\ge0.08$, recoil. After the launch transient, for radiation emitted after $\xi=20$, their rates at $N=1$ ($N\dtres$ from $0.04$ to $0.12$), $N=1.6$ and $N=2$ follow the same function, with median deviations between $-4.3\%$ and $+1.7\%$ per simulation; a larger deviation, a median of $-16\%$, occurs only for radiation emitted between $\xi=20$ and $30$ at $N\dtres=0.10$, shortly after the launch at strong dispersion.

The rate function closes a model of the recoil. A soliton that radiates the norm $-dQ$ at the wavenumber $k_r$ loses the momentum $-k_r\,dQ$, so that
\begin{equation}
\frac{dQ}{d\xi}=-RQ,\qquad \frac{dk_s}{d\xi}=-(k_r-k_s)\,R,
\label{eq:recoil}
\end{equation}
with $N=Q/(2\sqrt{D_2})$, $k_r-k_s=q_r(\tilde\delta)N/\sqrt{D_2}$ and $R$ from equation~\eqref{eq:ratefn}. Once that rate function is fixed, the recoil evolution introduces no additional fitting parameter. Initialized with the state of each launched simulation at $\xi=10$, it reproduces their subsequent evolution [figure~\ref{fig:emission}(b)--(d)]: $k_s$ to within $2.3\%$ at $\xi=20$ and $0.4\%$ at $\xi=160$, while the norm loss after $\xi=10$ that it predicts falls short of the simulated one by $9\%$ at $N\dtres=0.08$ and by $4$ to $5\%$ at $N\dtres=0.10$ to $0.12$ at $\xi=160$; the shortfall is the decay of the launch transient, which the model does not contain; by $\xi=160$ it makes up $81\%$ of the loss after $\xi=10$ at $N\dtres=0.06$ and $27\%$ at $N\dtres=0.07$. Along the recoil $k_s$ decreases, $D_2$ grows and $\tilde\delta$ falls, and with it the rate. Since $\ln R$ is steep in $\tilde\delta$, $R$ then decays as $1/(K\xi)$ with $K=d(1/R)/d\xi$ slowly varying, which is the algebraic law $-d\ln Q/d\xi\simeq c/\xi$ with $c\simeq1/K$; a slow decay of this kind, with a radiation-driven change of velocity, was found for the radiating solitons of the cubic third-order equation \cite{Karpman2001}, and here the measured rate function fixes it with no further fitting. The model reproduces the exponents extracted from the simulations over $\xi\in[40,80]$ and $[80,160]$ to within $2.5\%$ for $N\dtres\ge0.08$, including their drift: $c$ rises from $2.8\times10^{-3}$ to $4.3\times10^{-3}$ at $N\dtres=0.08$ and falls from $1.1\times10^{-2}$ to $1.0\times10^{-2}$ at $N\dtres=0.12$, towards $1/K$, which lies between $8.3\times10^{-3}$ and $9.8\times10^{-3}$ along the model in these windows [figure~\ref{fig:emission}(d)]. The momentum balance is also satisfied by the direct simulations: the measured $dk_s/d\xi$ equals $-(k_b-k_s)R$ with a median ratio between $0.98$ and $1.01$ for $N\dtres\ge0.08$. At the operating point the prepared soliton radiates $4.8\times10^{-9}$ of its norm per unit $\xi$, which is $3\times10^{-8}$ per surface-gravity period.

Two limits apply. The dispersion is truncated at third order, which matters at the resonant wavenumbers $k_r\simeq7$ to $14$ of table~\ref{tab:steady}, frequency offsets of $24$ to $46$~THz at $T_0=50$~fs. The rate function~\eqref{eq:ratefn} is an interpolation over its range of $\tilde\delta$, not an asymptotic law: the beyond-all-orders prefactor of the emission is measured here, not derived.

\subsection{Kinematic surface gravity: drive independence and the limit without slow light}
\label{sec:kappa}

\begin{figure}[tbp]
\centering\includegraphics{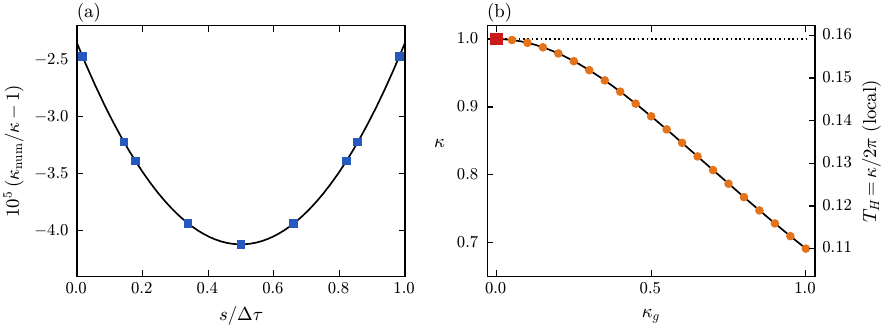}
\caption{Kinematic surface gravity at the operating point, equation~\eqref{eq:oppoint}. (a)~Relative deviation of the extracted $\kappa_{\rm num}$ from the closed form~\eqref{eq:kappaclosed} at nine drive amplitudes $d\in[0.7,1.3]$ (squares), plotted against the position $s/\Delta\tau$ of the crossing between grid nodes, with equation~\eqref{eq:extractbias} (line). (b)~$\kappa$ over $21$ values of $\kg\in[0,1]$ at $d=1$ (circles) and the closed form $N\,S(k_{\rm op})$ (line); square and dotted line: $\kappa=N$ at $\kg=0$. Right axis: the local kinematic surface-gravity temperature $T_H=\kappa/2\pi$ of the crossing.}
\label{fig:kinematics}
\end{figure}

Within the kinematic model~\eqref{eq:flow} the surface gravity of equation~\eqref{eq:kappa} does not depend on the drive; this is an identity of the model, not a numerical finding. Differentiating equation~\eqref{eq:flow} gives $|dv/d\tau|=\kappa_0\operatorname{sech}^2[\kappa_0(\tau-\tau_h)/\cmatch]$, and the crossing condition $v(\tau_h^{\times})=\cmatch$ places $\tau_h^{\times}=\tau_h$ at the inflection, where the $\operatorname{sech}^2$ factor is unity: a $\tanh$ profile has a translation-invariant central slope. With $\beta_0(k_{\rm op})=\tfrac12k_{\rm op}^2-\dtres k_{\rm op}^3$ in equation~\eqref{eq:slowlight}, equation~\eqref{eq:kappa} therefore has the closed form
\begin{equation}
\kappa \;=\; \kappa_0 \;=\; N\,S(k_{\rm op}) \;=\; \frac{N}{\sqrt{\,1+\big[\kg/\beta_0(k_{\rm op})\big]^{2}}},
\label{eq:kappaclosed}
\end{equation}
for every $d$. At the operating point~\eqref{eq:oppoint}, where $\beta_0(k_{\rm op})=0.95625$, it evaluates to $\kappa=0.954146$ and $T_H=\kappa/2\pi=0.151857$, the values used throughout except where the extracted $\kappa_{\rm num}$ is named. The finite-difference extraction specified after equation~\eqref{eq:kappa}, applied at nine drive amplitudes uniformly spaced over $d\in[0.7,1.3]$, yields
\begin{equation}
\kappa_{\rm num} = 0.954107, \quad \max_d \kappa_{\rm num} - \min_d \kappa_{\rm num} = 1.57\times10^{-5},
\label{eq:driveindep}
\end{equation}
a fractional spread of $1.6\times10^{-5}$ over a $\pm30\%$ variation of the drive, which moves the horizon by $\pm0.15$ [figure~\ref{fig:kinematics}(a)]. Both the spread and the offset of $4.1\times10^{-5}$ below equation~\eqref{eq:kappaclosed} are errors of the finite-difference extraction, which are available in closed form because the profile is known analytically: central differencing at the crossing and linear interpolation onto a crossing that falls a distance $s$ beyond the grid node below it give
\begin{equation}
\frac{\kappa_{\rm num}}{\kappa_0}\;=\;1-\lambda^{2}\!\left[\frac{\Delta\tau^{2}}{3}+s\,(\Delta\tau-s)\right]+\mathcal{O}(\Delta\tau^{4}),
\label{eq:extractbias}
\end{equation}
with $\lambda\equiv\kappa_0/\cmatch=N/|v_{g0}(k_{\rm op})|$ and $s\in[0,\Delta\tau)$ set by the horizon position and therefore by the drive (supplementary material, section S3). The deficit runs from $-\tfrac13(\lambda\Delta\tau)^{2}$ for a crossing on a node ($s=0$) to $-\tfrac{7}{12}(\lambda\Delta\tau)^{2}$ for a crossing midway between nodes ($s=\Delta\tau/2$), as the line of figure~\ref{fig:kinematics}(a) shows. At $d=1$ the crossing lies midway between nodes and the deficit is $-\tfrac{7}{12}(\lambda\Delta\tau)^{2}=-4.1185\times10^{-5}$, against $-4.1183\times10^{-5}$ from the extraction. For every $d$ the extracted $\kappa$ is confined to a band of absolute width $\tfrac14\kappa_0(\lambda\Delta\tau)^{2}=1.68\times10^{-5}$ that falls as $\Delta\tau^{2}$; the nine amplitudes sample this band, and equation~\eqref{eq:extractbias} reproduces all nine extracted values to $2.4\times10^{-9}$. The band bounds any drive dependence of the flow gradient on this grid; within the model there is none. Every conditional thermal prediction of the model, including the slope $-2\pi/\kappa$ of section~\ref{sec:thermality} and the squeezing parameter of section~\ref{sec:entanglement}, is fixed by $\kappa$ \emph{before} the drive is specified.

Figure~\ref{fig:kinematics}(b) shows the second property of the model: the horizon with slow light connects continuously to the one without it ($\kg\to0$ at fixed drive). Its right axis converts $\kappa$ into $T_H=\kappa/2\pi$, the local kinematic surface-gravity temperature of the crossing; no asymptotic emission at this temperature is implied. As the gap opens over $\kg\in[0,1]$, $\kappa$ decreases smoothly and monotonically from $0.99996$ to $0.69109$, with no threshold on the way, as equation~\eqref{eq:kappaclosed} requires. At the $\kg\to0$ endpoint the extraction reproduces the bare-soliton value $\kappa=N$ of the flow without slow light for three soliton amplitudes,
\begin{equation}
\kappa(\kg{=}0)\big|_{N=1,2,3} = \{0.99996,\;1.99967,\;2.99889\},
\label{eq:continuity}
\end{equation}
with residuals that grow as $N^{3}$ and that equation~\eqref{eq:extractbias} reproduces to better than one part in $10^{3}$ (supplementary material, section S3). At this endpoint the construction shares the dispersion model of reference~\cite{Oguz2026} but not its surface gravity, which there is the Kerr-index gradient at the velocity-matching point of a fissioning higher-order soliton, whereas here it is the gradient of the modeled flow of a stationary fundamental soliton.

\subsection{Emission channels and the imposed thermal relation}
\label{sec:thermality}

At a dispersive horizon of surface gravity $\kappa$, the near-horizon mode-mixing analysis predicts detailed balance between the Bogoliubov amplitudes of each Hawking pair \cite{Unruh1981,Brout1995,Coutant2012,Macher2009},
\begin{equation}
\left|\frac{\beta_\omega}{\alpha_\omega}\right|^2 = e^{-2\pi\omega/\kappa},
\qquad |\alpha_\omega|^2-|\beta_\omega|^2=1 .
\label{eq:thermal}
\end{equation}
Equation~\eqref{eq:thermal} is imposed as a model input, evaluated at the kinematic $\kappa$ of section~\ref{sec:kappa}; it is not derived here, and the scattering calculation of section~\ref{sec:discussion} does not reproduce it. All spectral statements refer to the analysis band $[\omega_{\rm lo},\omega_{\rm hi}]=[0.05,1.6]$, a parameter of the construction chosen independently of the operating point; the escape bound and the window shares below scale with its edges. The frequency $\omega$ is conjugate to $\xi$: it is an offset of the propagation constant, not an optical frequency. A long-wavelength mode of the positive-norm branch with $\omega>\mu$ has the wavenumber $k$ with $\tfrac12k^2+\mu-\dtres k^3=\omega$, $k\simeq\pm[2(\omega-\mu)]^{1/2}$, which is the optical detuning $k/T_0$ from the carrier (section~\ref{sec:model}); at the operating point the part of the band above $\mu$, $\omega\in[0.5,1.6]$, corresponds to $k$ from $-1.39$ to $1.62$, detunings from $-4.4$ to $5.2$~THz at $T_0=50$~fs. The pair amplitudes are generated from equation~\eqref{eq:thermal} at the extracted $\kappa_{\rm num}$ on $60$ frequencies of the band, with the Bogoliubov normalization enforced exactly (supplementary material, section S2). Whether a Hawking quantum of frequency $\omega$ can leave the horizon region at all is decided by the asymptotic channels, which do not depend on equation~\eqref{eq:thermal}.

In the asymptotic region the fluctuation branches reduce to $\Omega_\pm(k)=\pm(\tfrac12k^2+\mu)-\dtres k^3$, whose long-wavelength parts are gapped at $\mu=N^2/2$. The far-detuned pair of roots, near $|k|=k_{\rm RR}$, is the resonant-radiation lobe treated in section~\ref{sec:bdg}, and it lies outside the long-wavelength regime in which equation~\eqref{eq:thermal} is derived. The long-wavelength channels are evaluated with this dispersion relation, Doppler-shifted by the asymptotic flow. The measured drift $V_d$ of the soliton (section~\ref{sec:bdg}) enters in two ways, which agree at the operating point. The transformation to the frame comoving with the soliton, $\tau\to\tau-V_d\xi$, referred to below as the boost, Doppler-shifts both asymptotic branches, $\Omega\to\Omega-V_dk$; for the linear waves this change of frame is exact, although third-order dispersion leaves equation~\eqref{eq:gnlse} without Galilean symmetry. It lowers the long-wavelength escape threshold by $V_d^2/2$, up to corrections of order $\dtres|V_d|^3$ from the cubic term of the branch, and adds $-V_d$ to the downstream flow speed. The drift-adjusted matching speed $\cmatch'=|v_{g0}(k_{\rm op})-V_d|\,S(k_{\rm op})$ is instead a model-defined treatment, in which the drift rescales the flow speed downstream and leaves the upstream side at rest; the census below uses it. Under the drift-adjusted treatment, the pair channel, in which the negative-norm partner propagates downstream, is open on the part of the band that lies below the top of the Doppler-shifted negative-norm branch at low wavenumber,
\begin{equation}
\omega^{(-)}_{\max}=\max_{0<k<1/(3\dtres)}\big[\,2\cmatch' k-\tfrac12k^2-\mu-\dtres k^3\,\big],
\label{eq:admiss}
\end{equation}
which for $\dtres\to0$ and $V_d\to0$ reduces to the supersonic condition $2\cmatch>\sqrt{2\mu}=N$; under the boost the same expression holds with $2\cmatch'$ replaced by $2\cmatch-V_d$. This channel exists because the flow~\eqref{eq:flow} stays supersonic as $\tau\to\infty$; behind a physical soliton the flow returns to zero, so the downstream part of the census is a property of the kinematic model. Long-wavelength emission to the upstream side requires $\omega>\mu$ for a soliton at rest and under the drift-adjusted treatment. At the threshold $\omega=\mu$, the imposed relation~\eqref{eq:thermal} with $\kappa=N\,S(k_{\rm op})$ gives the occupancy
\begin{equation}
|\beta_\mu|^2=\frac{1}{e^{\pi N/S(k_{\rm op})}-1}\le\frac{1}{e^{\pi N}-1},
\label{eq:threshold}
\end{equation}
so the near-horizon occupancy at the threshold is controlled by $N/S$ alone, and emission inside the band requires $N<\sqrt{2\omega_{\rm hi}}\simeq1.79$, or $N^2-V_d^2<2\omega_{\rm hi}$ under the boost. Within the band, the limit is set by the chemical potential of the soliton, not by the flow.

\begin{table}[tbp]
\centering
\caption{Kinematic Hawking window at $\dtres=0.05$, $\kg=0.3$ with the drift measured at each $N$ over $\xi\in[30,40]$. Drift-adjusted criterion of equation~\eqref{eq:admiss} at $k_{\rm op}=1$ and $1.5$: upper edge $\omega^{(-)}_{\max}$ of the pair channel and the window $[\mu,\min(\omega^{(-)}_{\max},1.6)]$ with its share of the width of the band $[0.05,1.6]$. Boost to the comoving frame at $k_{\rm op}=1.5$: edge and share of the band width, with the escape threshold lowered by the drift. At $k_{\rm op}=1.5$, the exponent $\pi N/S$, the occupancy~\eqref{eq:threshold} and the negativity~\eqref{eq:EN} at $\omega=\mu$; for $\mu$ above the band ($N\ge1.8$), the occupancy only. The operating point is the row $N=1$ at $k_{\rm op}=1.5$.}
\label{tab:window}
\footnotesize\setlength{\tabcolsep}{4pt}
\begin{tabular}{llllllll}
\toprule
 & & \multicolumn{2}{l}{drift-adjusted} & boost & & & \\
$N$ & $\mu$ & $k_{\rm op}=1$ & $k_{\rm op}=1.5$ & $k_{\rm op}=1.5$ & $\pi N/S$ & $|\beta_\mu|^2$ & $E_N(\mu)$\\
\midrule
0.6 & 0.18 & $0.747$: $[0.18,0.747]$, 37\% & $1.963$: $[0.18,1.6]$, 92\% & $1.934$, 92\% & 1.976 & 0.161 & 0.782\\
0.8 & 0.32 & $0.638$: $[0.32,0.638]$, 21\% & $1.875$: $[0.32,1.6]$, 83\% & $1.821$, 83\% & 2.634 & 0.0773 & 0.549\\
1.0 & 0.50 & $0.497$: none & $1.760$: $[0.50,1.6]$, 71\% & $1.675$, 71\% & 3.293 & 0.0386 & 0.390\\
1.2 & 0.72 & $0.329$: none & $1.625$: $[0.72,1.6]$, 57\% & $1.499$, 50\% & 3.951 & 0.0196 & 0.279\\
1.4 & 0.98 & $0.146$: none & $1.491$: $[0.98,1.491]$, 33\% & $1.303$, 21\% & 4.610 & 0.0101 & 0.200\\
1.6 & 1.28 & $0.026$: none & $1.481$: $[1.28,1.481]$, 13\% & $1.148$, none & 5.268 & 0.00518 & 0.144\\
1.8 & 1.62 & $0.062$: none & $1.734$: none & $1.096$, none & 5.927 & 0.00267 & ---\\
2.0 & 2.00 & $0.209$: none & $2.161$: none & $1.094$, none & 6.585 & 0.00138 & ---\\
\bottomrule
\end{tabular}
\end{table}

\begin{figure}[tbp]
\centering\includegraphics{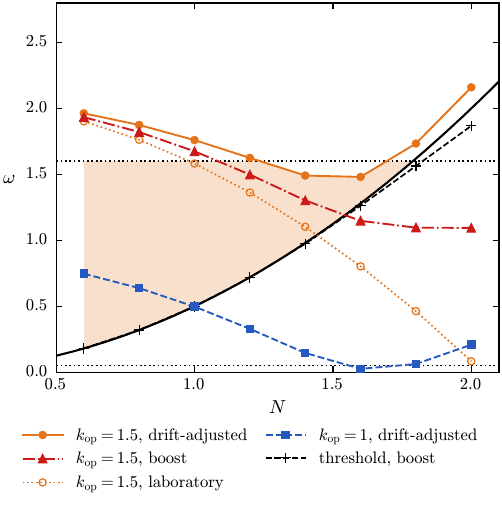}
\caption{Emission channels over the soliton amplitude at $\dtres=0.05$, $\kg=0.3$. Upper edge $\omega^{(-)}_{\max}$ of the pair channel, equation~\eqref{eq:admiss}, with the drift measured over $\xi\in[30,40]$: drift-adjusted criterion at $k_{\rm op}=1.5$ (filled circles, solid) and $k_{\rm op}=1$ (squares, dashed), boost to the comoving frame at $k_{\rm op}=1.5$ (triangles, dash-dotted), and laboratory frame at $k_{\rm op}=1.5$ (open circles, dotted); lines join the grid values. Solid black curve: the escape threshold $\omega=\mu=N^2/2$; plus signs joined by a dashed line: the threshold lowered by the drift under the boost; dotted lines: the band edges $\omega_{\rm lo}=0.05$ and $\omega_{\rm hi}=1.6$. Shaded: the kinematic Hawking window at $k_{\rm op}=1.5$ under the drift-adjusted criterion, between $\mu$ and $\min(\omega^{(-)}_{\max},1.6)$, with the edge interpolated linearly between grid values.}
\label{fig:channels}
\end{figure}

At the operating point~\eqref{eq:oppoint} the pair channel is open across the band from the start of the propagation: $2\cmatch=2.22$ exceeds $N=1$, the laboratory-frame edge $\omega^{(-)}_{\max}=1.58$ lies just below the top of the band, and with the drift-adjusted matching speed the edge lies between $1.73$ and $1.77$, above $\omega_{\rm hi}=1.6$, in each of the five propagation intervals over which the drift is measured, from $\xi\in[0,5]$ to $[30,40]$. Upstream escape is open above $\mu=0.5$, so the kinematic Hawking window $[\mu,\min(\omega^{(-)}_{\max},\omega_{\rm hi})]$ is $[0.5,1.6]$, $71\%$ of the width of the band, in every interval ($[0.5,1.58]$, $70\%$, in the laboratory frame), and at its lower edge equation~\eqref{eq:threshold} gives $|\beta_\mu|^2=0.0386$. The share is one of width, not of emission: with the imposed thermal relation~\eqref{eq:thermal} at $\kappa=0.954146$ the occupancy $|\beta_\omega|^2=(e^{2\pi\omega/\kappa}-1)^{-1}$ integrates to $0.193$ over the band and to $5.7\times10^{-3}$ over the window, so the window carries $3.0\%$ of the model-imposed pair flux of the band, the rest lying below $\mu$, where nothing escapes upstream. The boost to the comoving frame leaves this window unchanged: it yields an edge between $1.66$ and $1.68$ over the same intervals and lowers the escape threshold by $V_d^2/2\le1.51\times10^{-3}$ (supplementary material, section S6). Table~\ref{tab:window} and figure~\ref{fig:channels} give the window over the soliton amplitude, with the drift measured at each $N$. At larger amplitude the two treatments of the drift differ: under the boost the Hawking window at $k_{\rm op}=1.5$ covers $50\%$ and $21\%$ of the band at $N=1.2$ and $1.4$ and closes from $N=1.6$ on, where the drift-adjusted criterion yields windows of $57$, $33$ and $13\%$ of the band width, while the lowered threshold opens a narrow window, $[1.564,1.6]$, at $N=1.8$ for $k_{\rm op}\ge1.75$ (supplementary material, section S6). At $N=2$ and $k_{\rm op}=1.5$ the pair channel covers the whole band from $\xi\simeq10$ on under the drift-adjusted criterion of equation~\eqref{eq:admiss}, $67\%$ of it under the boost at the late drift $-0.53$, and $2\%$ in the laboratory frame. At $k_{\rm op}=1$ it is closed for the first $20$ units of $\xi$ and afterwards open only below $\omega\simeq0.11$ to $0.21$ under the drift-adjusted treatment, and it stays closed under the boost. The TOD term in equation~\eqref{eq:admiss} lowers the branch top well below its dispersionless estimate $2\cmatch'^{\,2}-\mu$, so a channel can remain closed after $2\cmatch'$ has exceeded $N$. At $N=2$ the escape threshold $\mu=2$ lies above the band for every $k_{\rm op}$, so no upstream emission is possible within the band.

Because equation~\eqref{eq:thermal} is imposed rather than derived, the corresponding log-linear relation provides only an internal consistency check and does not constitute an independent test of thermality: a regression of $\ln|\beta_\omega/\alpha_\omega|^2$ on $\omega$ returns the slope $-2\pi/\kappa_{\rm num}$ with which the amplitudes were generated (supplementary material, section S2). The slope $-2\pi/\kappa$ is therefore fixed by the flow gradient of the kinematic model, which the soliton amplitude and the slow-light factor set before the drive is specified.

An observer would collect the transmitted flux, not the bare ratio. The soliton itself does not filter it in the integrable limit: the linearization of equation~\eqref{eq:gnlse} about $\psi_0$ is reflectionless, and at $N=1$ its continuum eigenfunctions are exactly
\begin{equation}
\begin{aligned}
u_k&=e^{ik\tau}(k+i\tanh\tau)^2,\qquad v_k=e^{ik\tau}\operatorname{sech}^2\tau,\\
\Omega&=\tfrac12(1+k^2),
\end{aligned}
\label{eq:greybody}
\end{equation}
whose two asymptotes differ only by the unimodular factor $(k+i)^2/(k-i)^2$ \cite{Kaup1990}. The transmission through the soliton at rest is therefore unity for $\Omega>\mu$, which at the operating point is the upper $71\%$ of the width of the band, while for $\Omega<\mu$, the lower $29\%$, no propagating long-wavelength state exists. Grey-body filtering can therefore arise only from integrability breaking or from the flow. A mode that escapes upstream must cross the flow~\eqref{eq:flow}. In the linear operator of the model, which adds to the BdG operator of equation~\eqref{eq:bdg} the advection $F=-\tfrac{i}{2}(v\partial_\tau+\partial_\tau v)$ by the flow (supplementary material, section S7), the substitution $u=e^{-i\int v\,d\tau}w$ on the positive-norm branch removes $F$ and leaves the potential step $-\tfrac12v^2$, which reflects the escaping wave and suppresses its transmission as $\omega\to\mu^+$. In that calculation, at $(N,k_{\rm op})=(1,1.5)$ and $\dtres=0$, the upstream population falls to $0.14$ of the Planck form at the lowest band frequency above $\mu$ [supplementary material, figure~S1(b)]. The values of equation~\eqref{eq:threshold} at the threshold are therefore near-horizon occupancies and not fluxes at a detector; the grey-body factor of the physical problem is not computed here. The slope $-2\pi/\kappa$ still depends on $\kg$ through the flow of figure~\ref{fig:kinematics}(b).

\subsection{Idealized two-mode squeezing and entanglement}
\label{sec:entanglement}

\begin{figure}[tbp]
\centering\includegraphics{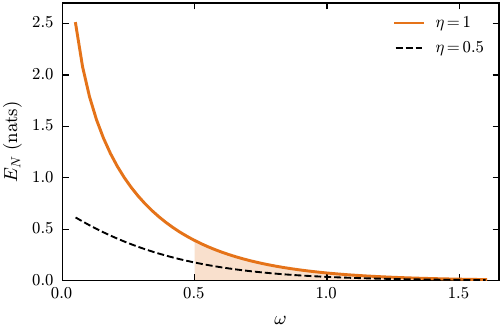}
\caption{Logarithmic negativity $E_N(\omega)=2\,\operatorname{arcsinh}|\beta_\omega|$ of the idealized near-horizon pair state with the imposed Bogoliubov amplitudes of equation~\eqref{eq:thermal} at the operating point ($\kappa=0.954146$; solid), and after symmetric pure loss of transmissivity $\eta=0.5$, equation~\eqref{eq:loss} (dashed). Shaded: the Hawking window $\omega\in[\mu,1.6]$.}
\label{fig:EN}
\end{figure}

Classical noise can mimic a thermal spectrum but not the nonseparability of the Hawking mode and its partner, the negative-norm mode on the far side of the crossing, which theory identifies as the signature of spontaneous pair production \cite{Finazzi2014,Busch2014} and a condensate experiment measured \cite{Steinhauer2016}; it has been analyzed for stimulated Hawking radiation in an optical white-hole/black-hole pair \cite{Agullo2022}, and the single-photon experiment of reference~\cite{FelipeElizarraras2026} provides a platform for exploring it in fibers. Within the model the diagnostic has a closed form, because the linear dynamics is Gaussian: the pair Bogoliubov transformation with the imposed amplitudes $(\alpha_\omega,\beta_\omega)$ acting on the vacuum (covariance matrix $\sigma_{\rm vac}=\openone$ in our convention) produces at each frequency a two-mode squeezed vacuum with squeezing parameter $r_\omega=\operatorname{arcsinh}|\beta_\omega|$. The smaller symplectic eigenvalue of its partially transposed covariance matrix is $\numinus=e^{-2r_\omega}$, the state is entangled iff $\numinus<1$ \cite{Simon2000}, and the logarithmic negativity is \cite{Vidal2002,Plenio2005,Adesso2007}
\begin{equation}
E_N(\omega) = -\ln\numinus = 2\,\operatorname{arcsinh}|\beta_\omega|
= 2\ln\!\big(|\alpha_\omega|+|\beta_\omega|\big),
\label{eq:EN}
\end{equation}
the last form by the Bogoliubov normalization in equation~\eqref{eq:thermal}. This pair state is an idealization of the model: above $\mu$ three asymptotic channels are open in the model's own scattering (supplementary material, section S7), so the reduced state of the Hawking mode and its partner is in general mixed, and below $\mu$ no upstream channel exists. The values below therefore refer to the idealized two-mode pair at an independently fixed $\kappa$; the negativity that Hawking radiation generates for a lattice-regularized field on a $1+1$-dimensional analogue black-hole background is a different quantity, with a UV-finite volume term that encodes the number density of entangled pairs \cite{ChandranFischer2026}. Since equation~\eqref{eq:thermal} gives $|\beta_\omega|>0$ at every $\omega>0$, $\numinus<1$ throughout, and only the magnitude of $E_N$ carries information. The resulting logarithmic negativity characterizes the idealized two-mode state implied by the imposed Bogoliubov amplitudes; it is not a dynamical prediction, either of the soliton dynamics or of the scattering calculation (supplementary material, section S7).

Figure~\ref{fig:EN} shows $E_N$ at the operating point. On the Hawking window $[\mu,1.6]$ it falls from $0.390$ nats at $\omega=\mu$ to $0.0103$ at $\omega=1.6$. Below $\mu$ the state is a near-horizon pair state and not a flux at a detector; its negativity rises to $2.50$ nats at $\omega=0.05$ along the infrared law $E_N\to\ln(2\kappa/\pi\omega)$, so that its value at the lower band edge is set by that edge (halving $\omega_{\rm lo}$ adds $\ln2\simeq0.69$ nats). Because $|\beta_\omega|$ depends on $\omega/\kappa$ alone, the curve inherits the drive independence of section~\ref{sec:kappa}. Experimentally, $E_N$ of a two-mode Gaussian state is fixed by second moments, so quadrature (homodyne) correlations between the Hawking and partner channels suffice \cite{Adesso2007,Finazzi2014,Busch2014}. Classical phase-insensitive noise cannot drive $\numinus$ below unity, but parametric processes of the pump, such as spontaneous four-wave mixing, produce nonseparable pairs without a horizon, so nonseparability alone does not identify Hawking emission. A symmetric pure-loss channel of transmissivity $\eta$ acting on both modes maps the smaller partially transposed symplectic eigenvalue to
\begin{equation}
\numinus'=1-\eta\,\bigl(1-e^{-2r_\omega}\bigr),
\label{eq:loss}
\end{equation}
which stays below unity for every $\eta>0$: at $\omega=\mu$, $E_N=-\ln\numinus'$ is $0.390$, $0.344$, $0.176$ and $0.033$ nats at $\eta=1$, $0.9$, $0.5$ and $0.1$. Phase-insensitive excess noise $\bar n$ in both input modes shifts the negativity to $\max\{0,\,E_N-\ln(1+2\bar n)\}$, so nonseparability survives up to $\bar n=0.24$ at $\omega=\mu$ and up to $\bar n=5.2\times10^{-3}$ across the whole window; the thermal occupation of optical modes, about $4\times10^{-14}$ at $1550$~nm and $300$~K, is negligible against either (supplementary material, section S9).

\section{Discussion}
\label{sec:discussion}

\emph{Status of the results.} Derived from equation~\eqref{eq:gnlse}, without the horizon model, are the BdG spectrum about the stationary periodic-box TOD state, the spurious quartet about the integrable profile and the radiative loss of launched solitons (section~\ref{sec:bdg}), the phase matching and the steady rate of the Cherenkov emission and the recoil that norm and momentum balance imply (section~\ref{sec:emission}), and the asymptotic channel structure, whose escape condition $\omega>\mu$ does not use the flow model and whose pair channel uses only its downstream speed (section~\ref{sec:thermality}). Empirically parameterized is the rate function~\eqref{eq:ratefn}, whose three parameters are fitted to the prepared solitons; the recoil evolution adds none. Imposed by the kinematic model are the thermal relation~\eqref{eq:thermal}, the drive-translated horizon~\eqref{eq:tauh}, the phenomenological slow-light factor~\eqref{eq:slowlight} and the idealized two-mode state with its logarithmic negativity~\eqref{eq:EN}. Not demonstrated are dynamical Hawking thermality, dynamically generated pair entanglement and a fully driven stationary soliton-horizon solution.

\emph{Relation to the undriven theory.} Reference~\cite{Oguz2026} introduced the coarse-grained thermodynamics of solitonic horizons in free flight. This paper keeps its dispersion model and adds the linear-response, emission and channel analysis of the soliton that supports the horizon, together with a kinematic model of the flow in which thermality and entanglement are evaluated at an independently fixed $\kappa$. The $\kg$ dependence of figure~\ref{fig:kinematics}(b) reaches the $\kg=0$ endpoint~\eqref{eq:continuity}, and the Cherenkov resonance~\eqref{eq:resonance}, phase-matched in the frame of the drifting soliton, sets the radiative channel of sections~\ref{sec:bdg} and~\ref{sec:emission}. No statement about the effect of a drive or of slow light on the generalized second law of reference~\cite{Oguz2026} is made here.

\emph{Model inputs.} The near-horizon thermal form~\eqref{eq:thermal} is \emph{evaluated} at the independent kinematic $\kappa$ and constrained by the exact Bogoliubov normalization. A direct scattering calculation at $\dtres=0$ for an operator assembled from the BdG operator of the soliton and the advection by the model flow (supplementary material, section S7), which is not the linearization of a flowing stationary state, does not reproduce it. At $(N,k_{\rm op})=(2,3)$, where only the downstream channels are open in the band, the conversion ratio $|S_{Rv\leftarrow Ru}|^2/|S_{Ru\leftarrow Ru}|^2$ stays between $0.48$ and $0.50$ for $\omega\in[0.2,1.6]$ while the Boltzmann factor falls by a factor of $82$; at $(1,1.5)$, where upstream escape is open above $\omega=\mu$, the upstream occupancy departs from the Planck form at the kinematic $\kappa$ of that point by factors between $0.14$ and $17$. The reason is that in this operator the $u$--$v$ coupling enters only through $\psi_0^2$, which is confined to a width $1/N$ and does not depend on the flow gradient, while the flow enters only through the advection, so $\kappa_0$ cannot set the mode conversion. This negative result delimits the scope of the kinematic construction: equation~\eqref{eq:thermal} remains a model input that the construction does not derive, and the channel structure of section~\ref{sec:thermality} confines it to a near-horizon role. The calculation is a statement about this operator, not about solitonic horizons, and it does not test the dependence of the spectrum of a dispersive index front on the front: there the gradient of the front sets the temperature, with a frequency-dependent grey-body factor \cite{Linder2016}, and the profile of the front and the dispersion correct the thermal spectrum \cite{Jacquet2015,Bermudez2016}, and a non-monotonic dispersion adds negative-energy branches that compete with the Hawking channel \cite{HolandaRibeiro2023}, whereas in this operator the flow gradient does not enter the mode conversion at all. The flow profile~\eqref{eq:flow} is a kinematic ansatz whose defining property, that the drive enters only through the horizon position $\tau_h$, is the hypothesis that the dispersion-computed $v_g$ must reproduce. The slow-light factor~\eqref{eq:slowlight} is a phenomenological two-band envelope applied to the background flow: it rescales $\kappa_0$ and $\cmatch$ in equation~\eqref{eq:flow} and does not modify the operator $M$ of equation~\eqref{eq:bdg}, which contains no grating term and no grating-induced dispersion curvature, so the spectra of section~\ref{sec:bdg} are independent of $\kg$ and $k_{\rm op}$ by construction. A Floquet treatment of the periodic index modulation, with the grating inside the linearized operator and the spectrum read off the resulting band structure, lies outside the scope of this paper; it would change the radiative loss but not the local horizon kinematics, which enter only through the group velocity at the crossing and are fixed there by equation~\eqref{eq:kappaclosed}. The integrable continuum~\eqref{eq:greybody} is reflectionless for the soliton at rest; grey-body filtering by the flow and by integrability breaking is not computed here. The occupancy~\eqref{eq:threshold} at the threshold uses the flow model through $\kappa$. All results refer to the same point: at $N=1$, $\dtres=0.05$ the launched background loses $9.3\times10^{-4}$ of its norm over the simulated interval corresponding to $6.1$ surface-gravity periods, almost all of it the shedding of its launch mismatch, while the prepared soliton radiates $4.8\times10^{-9}$ per unit $\xi$, and the BdG spectra of section~\ref{sec:bdg} are evaluated at the same $N\dtres=0.05$, where the standing tail of the periodic stationary TOD state is $3.6\times10^{-5}$.

\emph{Experimental outlook.} The normalization of section~\ref{sec:model} provides the physical-units dictionary (anomalous dispersion $\beta_2<0$, $T_0\sim50$~fs pulses, engineered $\beta_3$ \cite{Yuksel2025}) \cite{Agrawal2013,Dudley2006}. For the fiber parameters of reference~\cite{Oguz2026} ($\beta_2=-15~\mathrm{ps^2/km}$, $\gamma=0.1~\mathrm{W^{-1}km^{-1}}$, $T_0=50$~fs) the operating point~\eqref{eq:oppoint} corresponds to a dispersion length $L_D\simeq17$~cm, soliton peak power $P_0\simeq60$~kW (pulse energy $2P_0T_0/N\simeq6$~nJ), an engineered $\beta_3=6|\beta_2|T_0\,\dtres\simeq0.23~\mathrm{ps^3/km}$, and an operating offset $k_{\rm op}/(2\pi T_0)\approx4.8$~THz on the blue side of the carrier. In silica at $T_0=50$~fs, however, intrapulse Raman scattering would dominate the soliton kinematics on which the construction relies: at $N=1$ the spectral centroid $k_s$ of the soliton starts to shift at Gordon's rate $|dk_s/d\xi|=\tfrac{8}{15}(T_R/T_0)N^4\simeq0.03$ \cite{Gordon1986,Mitschke1986}. An integration of equation~\eqref{eq:gnlse} with a Raman response drives the soliton velocity from the TOD drift $-0.05$ to between $-0.6$ and $-0.9$ by $\xi=40$, while self-steepening alone ($1/\omega_0T_0\simeq0.016$) offsets the drift by $-0.017$ (supplementary material, section S8). The construction therefore maps onto a Raman-free Kerr medium, such as a noble-gas-filled hollow-core fiber, in which the Kerr response of the monatomic gas is instantaneous and the dispersion is pressure-tunable \cite{Russell2014}. The nonlinear coefficient of such a fiber is far below the value used above, so the soliton power $P_0\propto|\beta_2|/(\gamma T_0^2)$ and the pulse energy rise in proportion. Slow light from a Bragg structure has not been demonstrated in such fibers, and whether one gas, pressure and core realize the dictionary together, with the peak intensity below the photoionization threshold, is left open; a silica realization of the construction requires the Raman term in equation~\eqref{eq:gnlse}. The two-band factor~\eqref{eq:slowlight} has no fixed mapping onto a specific grating geometry; a coupled-mode treatment of a Bragg structure \cite{Mok2006} must supply the physical value of $\kg$. Within the model the two parameters act separately: $\kg$ sets $\kappa$ through $S$ [figure~\ref{fig:kinematics}(b)], while the drive only translates the horizon [figure~\ref{fig:kinematics}(a)], so the conditional thermal slope and the idealized entanglement spectrum would be controlled by $\kg$ at fixed drive. This separation is a property of equation~\eqref{eq:flow} until a coupled-mode calculation fixes $\kg$ and a driven model tests equation~\eqref{eq:tauh}. Neither removing the slow-light structure nor removing the drive switches the horizon off: at $\kg=0$ the model gives $\kappa=N$, equation~\eqref{eq:continuity}, and undriven solitons form horizons in fibers \cite{Philbin2008,Drori2019,Oguz2026}. A control for classical spectral broadening, the standard confounder in optical analogue experiments, must instead remove the group-velocity crossing itself, for example by detuning the probe out of the range of velocities that the soliton matches. The entanglement itself is a second-moment observable, equation~\eqref{eq:EN}, accessible to homodyne correlation measurements between the Hawking and partner bands \cite{Adesso2007,Finazzi2014,Busch2014}; the nonseparability threshold $\numinus<1$ gives a necessary but not sufficient target, insensitive to the drive, of the kind used in the condensate experiments \cite{Steinhauer2016,deNova2019,Kolobov2021}. The recoil-limited regime of the loss in sections~\ref{sec:bdg} and~\ref{sec:emission} is the classical backreaction of the emitted Cherenkov radiation on the soliton, which equation~\eqref{eq:recoil} describes. The backreaction of quantum fluctuations on optical solitons \cite{BaakKoenig2025} and that of stimulated Hawking radiation in an optical analogue \cite{Procopio2026} lie beyond the linear treatment used here \cite{Fischer2022_Backreaction,Fischer2024_Nonlinear}. Pumped solitons that emit dispersive waves are well established in microresonators, where the Lugiato--Lefever equation fixes the soliton width through the pump detuning and the dispersive wave is measured directly \cite{Coen2013,Brasch2016}. That setting supplies the explicit drive absent from equation~\eqref{eq:gnlse}, and there the hypothesis behind equation~\eqref{eq:tauh}, that the pump moves the horizon without reshaping the soliton, can be tested.

\section{Conclusions}
\label{sec:conclusions}

Three properties of solitons with third-order dispersion that underlie optical analogue horizons are obtained without any assumption about the emission from the horizon; the first is a statement about a periodic grid, and the third uses the model flow for its downstream speed and a fixed analysis band. The linearized problem must be posed about the stationary TOD state on the periodic computational domain, where the zero-mode sector consists of two $2\times2$ Jordan blocks at the origin; linearizing about the integrable profile produces a converged but spurious quartet scaling as $\sqrt{\dtres}$. An adiabatically prepared soliton settles to a steady emission rate, measured from the flux of the radiation that it leaves behind between $2.8\times10^{-14}$ and $8.2\times10^{-6}$ per unit $\xi$ for $\dtres$ from $0.035$ to $0.07$; the radiation is phase-matched in the frame of the drifting soliton, and the first-order rate falls short of the measured one by factors of $29$ to $45$, a beyond-all-orders prefactor that the measurement supplies. Expressed through the scaled dispersion coefficient of the local soliton, the rate describes launched solitons of three amplitudes; with norm and momentum balance, and no further fitted parameter, it reproduces their recoil and the algebraic decay of their loss rate, including the values of its exponent. At weak dispersion the loss of a launched soliton is the shedding of its launch mismatch: at the operating point, $N=1$ and $\dtres=0.05$, the launched soliton loses $9.3\times10^{-4}$ of its norm by $\xi=40$, while the prepared one radiates $4.8\times10^{-9}$ per unit $\xi$, so the background is quasi-static. Long-wavelength emission to the upstream side requires the chemical potential of the soliton, lowered by $V_d^2/2$ in the frame comoving with a soliton drifting at $V_d$ (unchanged under the drift-adjusted treatment), to lie inside the analysis band, and at the operating point the kinematic Hawking window covers $71\%$ of the width of the band but only $3\%$ of the pair flux that the imposed thermal relation assigns to it, whether the drift is applied as a boost to the comoving frame or through the drift-adjusted matching speed. In the kinematic model the surface gravity is $\kappa=N\,S(k_{\rm op})$ in closed form, with no dependence on the drive, and the extracted value varies over a $\pm30\%$ drive range by $1.6\times10^{-5}$, inside a closed-form discretization band. The thermal relation and the logarithmic negativity of the idealized two-mode state built on it, $E_N$ from $0.390$ nats at the escape threshold to $0.0103$ at the top of the band, are model inputs evaluated at that $\kappa$, not predictions of the dynamics. A scattering calculation for an operator assembled from the soliton and the model flow does not reproduce the thermal relation, so the analogue-Hawking interpretation remains conditional on the model; deriving the emission spectrum from the dynamics of a driven soliton in a Raman-free Kerr medium remains open.

\section*{Acknowledgments}
The author thanks U.~R.~Fischer for helpful discussions. The author gratefully acknowledges the support of the Material Physics Simulation Laboratory at Pamukkale University. This work was supported by the Istanbul Okan University Scientific Research Projects Coordination Unit (IOU BAP) under Project No.\ OBAP2026010006, and by the Pamukkale University Scientific Research Projects Coordination Unit (PAU BAP) under Project No.\ 2025ALDEP037. Computing resources were provided by the National Center for High Performance Computing of Turkey (UHeM) under grant number 5027772026.

\section*{Data availability}
The data that support the findings of this study are openly available at the following DOI: \url{https://doi.org/10.5281/zenodo.23156910} \cite{P2Data}.

\end{document}